\documentclass[onecolumn,aps,preprint,showpacs,amsmath,amssymb]{revtex4}
\usepackage{graphicx}
\usepackage{latexsym}
\usepackage{dcolumn} 
\usepackage{stmaryrd} 
 
\usepackage{bm}

\begin{document}
\newcommand{\eq}{\begin{equation}}                                                                         
\newcommand{\eqe}{\end{equation}}             
 
\title{Interaction of antiproton with helium based on ab-initio calculations  }

\author{Imre F. Barna$^{1,2}$, Mih\'aly A. Pocsai$^{1,3}$ and     
K. T\H{o}k\'esi$^{2,4}$}
\address{ $^1$ Wigner Research Center for Physics of the Hungarian Academy of Sciences  Konkoly-Thege Mikl\'os \'ut 29 - 33, 1121 Budapest, Hungary 
$^2$  ELI-HU Nonprofit Kft.,  Dugonics T\'er 13, H-6720 Szeged, Hungary 
$^3$ University of P\'ecs, Institute of Physics, Ifj\'us\'ag \'utja 6 H-7624 P\'ecs, Hungary,  
$^4$ Institute of Nuclear Research of the Hungarian Academy of Sciences 
(ATOMKI), H-4001 Debrecen, P.O. Box 51, Hungary  }
\date{\today}

\date{\today}

\begin{abstract} 
   We present ionization cross sections for antiproton and helium collisions based on {\it{ab-initio}} time-dependent coupled channel method. In our calculations a finite basis set of regular helium Coulomb wave packets and Slater function were used. The semiclassical approximation was applied with the time-dependent Coulomb potential to describe the antiproton electron interaction. Three different projectile energies were considered as 10, 50 and 100 keV. We found clear evidence for the formation of the anti-cusp in the differential distributions. 
\end{abstract}

\maketitle

 
\section{Introduction}
\label{introduction}
Collision between antiprotons and atoms has a fundamental
interest in atomic physics. In the nineties numerous experimental and theoretical works were done with low energy antiprotons 
investigating various interesting phenomena \cite{conf}.  
Protons and antiprotons are the lightest heavy ions, where the semiclassical approximation is valid, therefore 
their motion can be treated classically. 
Contrast to protons at antiproton collisions---due to the negative charge---no electron transfer can take place which makes 
the electron dynamics much simpler at low energies.  

In theoretical multi-electron atomic physics the investigation of antiproton helium collision has crucial interest. 

There are three existing benchmark experiments where total cross ionization sections were measured for low energy collisions between antiprotons   
and helium atoms.  The first two measurements were performed in 1990 \cite{and} and 1994 \cite{hve}. More recently, single and double ionization cross sections of He and Ar atoms were investigated \cite{hve2} by antiproton impact.  

These experiments induced  the  ``competition'' between theories for  the better and better descriptions of the physical processes, which is still very much active. In general, all  theoretical descriptions predict almost the similar results  above 100 keV antiproton collision energy, due to the weakening Sommerfeld parameter. This means that the interaction becomes perturbative and the first Born approximation is valid.   
The deviation between theory and experimental data becomes clearly visible between 10 and 100 keV impact energy and below 10 keV projectile energy, there are large deviations between non-perturbative {\it{}ab-inito} calculations. This is a clear indicator that the role of electron-electron correlation is crucial.  However, until now, there has been no clear evidence of which theoretical model is  superior.  \\
This paper now starts with a brief and non-exhaustive historical overview of the relevant existing theoretical models.  
The forced impulse method (FIM), developed by Reading and Ford \cite{red} was one of the most successfully early method for proton  and antiproton-helium collisions. An improved version of FIM is the multi-cut forced impulse (MFIM) method by \cite{red2}.
Bent {\it{et al.}} \cite{bent} used  multi-electron hidden crossing (MEHC) theory in their study of the single ionization of He in antiproton impact. 
The time-dependent density functional theory \cite{lud} is another powerful method to describe non-perturbative many-electron ionization processes, even
in the low keV/amu impact energy range. Keim {\it{et al.}} \cite{keim} applied density functional theory with various response functions and with a basis representation obtained from the basis generator 
method (BGM) to obtain single- and double-ionization cross sections in antiproton and helium collision systems. Later, the method was improved and the same scientific question was subsequently revised \cite{keim2}. 
Single ionization of He antiproton impact was also studied by Tong {\it{et al.}} \cite{tong} using a self-interaction-free time-dependent density-functional theory (SIF-TDDFT).

Various independent particle close-coupling methods in a semi-classical
impact parameter treatment, where the electron wave function is expanded
around the target nucleus, are prominent among the theoretical studies \cite{schi,wherm,igar,diaz,lee,pind}.

Later, a fully correlated, three dimensional approach was developed by Schulz and Krstic \cite{schulz} to study the ionization cross sections in collision between antiproton and helium atoms. They solved the time-dependent Schr\"odinger equation (TDSE) in a four dimensional Cartesian lattice (LTDSE) and calculated the 
ionization cross sections using $75^4$ lattice points.
The B-spline basis for the construction of the active electron wave function was used by Sahoo {\it{et al.}} \cite{saho}. B-splines have been widely
used in atomic physics\cite{mart} because of their ability to accurately represent the continuum channels when compared with other
conventional bases. Foster \cite{foster} published calculations obtained from lattice time dependent close coupling method. 

Fainstein {\it{et al.}} \cite{fai} applied the Coulomb Distorted Wave Eikonal 
Initial State (CDW-EIS) method below 100 keV antiproton impact energies and had an  
astonishingly good agreement with experimental data of \cite{hve}. This induced 
a debate regarding the validity of CDW-EIS method.

The independent particle approach also has been performed by
Schultz \cite{schu2} using a classical trajectory Monte Carlo Method (CTMC).

First we applied a time dependent coupled channel (TDCC) method in a special basis for calculating the single
ionization cross sections. Our results for angular differential ionization cross sections were compared with the results of CDW-EIS and CTMC methods  \cite{barna0}. Later also the energy and angular differential electron
emission cross sections were presented using CDW-EIS and CTMC models \cite{tok}.

In 2011,  Kirchner and Knudsen published a topical review which exhaustively discussed and compared all the experimental results and theoretical model calculations \cite{kirch2}.

In parallel to the development of theoretical ab-initio
methods, there must also be a focus on the deeper understanding of the dynamics of the anti-cusp dip, where the
DDCSs have to be analyzed.
	
 One of the intersting part of the collision physics is the study of electron emissions in the direction of the projectile motion. The electron capture to continuum (ECC) peak was discovered in 1970 by \cite{rudd} in the  double-differential cross section 
(DDCS) of electrons ejected in proton-atom collisions at 0 degree compared to the projectile initial velocity. 

For positively charged particles, the production of the cusp peak is experimentally and theoretically understood \cite{stol}.
Numerous investigations lead to the conclusion that the cusp is generated when the asymptotic velocity of the ionized electron 
is equal with the velocity of the projectile.

It can be explained for positively charged projectiles as result of the special case of ionization, where the ionized target electron is strongly influenced by the outgoing projectile, or in other words when the atomic electron is dragged by the projectile continuum state and moves with it. Following this scenario, using the negatively charged projectiles we can expect electron yield deficit in the direction of the projectile path and we can expect anti-cusp formation instead of cusp formation.

However, for negative charged projectile impact, due to the very limited available experimental data, the production of the anti-cusp
is barely understood, especially at low impact energies.
The anti-cusp supposes that a well defined gap must exist in the energy spectrum of the ionized electron in the direction of the projectile. Numerous theoretical studies have addressed the question of anti-cusp from different authors \cite{reinh1,yam,burg,fain2,ols2,reinh3}. To the best of our knowledge, only the total ionization cross section data is so-far available from experiments and there are no available {\it{ab initio}} calculations for differential ionization cross sections for antiproton helium 
impact. 

In this work we present ionization cross sections for antiproton and helium collisions based on {\it{ab-initio}} time-dependent coupled channel method. In our calculations a finite basis set of regular helium Coulomb wave packets and Slater functions are used. We show results for singly- and doubly differential cross sections at  10, 50 and 100 keV antiproton impact energies. So we believe, that this study might stimulate experimentalists to investigate this problem in the low energy antiproton-atom collisions in the future.


This paper contains a brief outline the used theoretical approach (Section 2) with the calculated results  are presented and 
discussed in Section 3. 
Atomic units are used throughout the paper unless otherwise indicated. 

\section{Theory}
\label{theory}
The TDCC method has been widely used in various
fields of atomic collision physics with the recognition that it is one of the most reliable and
powerful theoretical approaches \cite{kirch2}. Our single-center coupled-channel method was first introduced 
for the ionization of helium in relativistic heavy ion collisions \cite{imre,imre2}. 
Later, the same model was applied for positron helium collisions \cite{imre3}; for 
the photoionization \cite{imre4} and finally for two-photon double-ionization of helium \cite{imre5}. Currently, a very similar 
model is used to investigate photoionization of rubidium atoms \cite{Misi}.

In the semi-classical approximation, the projectile moves on a straight-line trajectory,
with constant velocity $v$ and impact parameter $b$.
The projectiles are considered to be classical point
charges without any inner structure.

To study the ionization process,   the time-dependent Schr\"odinger equation is solved with a time-dependent external Coulomb field:
\eq
i\frac{\partial }{\partial t}\Psi({\bf{r}}_1,{\bf{r}}_2,t) =
\left(\hat{H}_{He}+ \hat{V}(t)  \right)\Psi({\bf{r}}_1,{\bf{r}}_2,t),
\label{Sch:time}
\eqe
where $\hat{H}_{He}$ is the Hamiltonian of the unperturbed
helium atom
\eq
\hat{H}_{He}= \frac{p_1^2}{2} + \frac{p_2^2}{2} - \frac{2}{r_1}  -\frac{2}{r_2}
+ \frac{1}{|\bf{r}_1 -\bf{r}_2|}.
\eqe
The external time-dependent field $\hat{V}(t)$ is the antiproton-electron interaction
 \eq
 \hat{V}(t) = -\left(\frac{1}{R_1(t)} +\frac{1}{R_2(t)}  \right)
 \eqe
 with $R_i(t) = ((x_i-b)^2 + y_i^2 + (z_i-v_p t)^2)^{1/2}, i=1,2 $.
 Below 20 keV, impact energies instead of the straight-line trajectories Coulomb hyperbolas 
 are used. For collision energies below 1keV, the trajectory bending effect  is 
  even more important.  Originally the retarded Lenard-Wi\'echert potentials 
were used for relativistic collisions \cite{imre}.    
 Eq. (\ref{Sch:time}) is solved by the expansion of  $ \Psi({\bf{r}}_1,{\bf{r}}_2) $ in the basis
 of eigenfunctions $\{\Phi_j\}$ of the time-independent Schr\"odinger equation:
\eq
\hat H_{He} \Phi_j({\bf{r}}_1,{\bf{r}}_2) = E_j \Phi_j({\bf{r}}_1,{\bf{r}}_2),
\label{TISE}
\eqe
with 
\eq
\Psi({\bf{r}}_1,{\bf{r}}_2,t) =\sum\limits_{j=1}^{N}a_j(t)
 \Phi_j({\bf{r}}_1,{\bf{r}}_2)e^{-iE_jt} \label{tim-w},
\eqe
where $a_j(t)$ are the time-dependent expansion coefficients for the various
channels described by the wave functions $\Phi_j$. Inserting this Ansatz 
 into Eq.
 (\ref{Sch:time}) leads to a system of first-order
differential equations for the expansion coefficients:
\eq
\frac{da_k(t)}{dt}=-i\sum\limits_{j=1}^{N}V_{kj}(t)a_j(t) e^{i(E_k-E_j)t},\quad
(k=1,\dots,N),
\label{mivan}
\eqe
where $V_{kj}$ is the coupling matrix
$\langle  \Phi_k({\bf{r}}_1,{\bf{r}}_2) |\hat{V}(t)|
\Phi_j({\bf{r}}_1,{\bf{r}}_2) \rangle $
including the symmetrized products of
the projectile-elec-tron single-particle interaction matrix elements and
electron-electron single-particle overlap matrix elements, respectively.

Denoting the ground state with $k=1$, the following initial
conditions are used
\eq
 \begin{aligned}
 a_{k} \left( t \rightarrow - \infty \right) = 
 \end{aligned}
 \left\{
 \begin{aligned}
 & 1\quad & \text{for k}  = 1,   \\
 & 0\quad & \text{for k}  \neq 1.  
 \end{aligned}
 \right.
\eqe

The total cross section for occupying the helium eigenstate $k$
can be calculated as
 \eq
\sigma_k = \int P_k({\bf{b}},t\rightarrow \infty) d^2b, 
\eqe
with the probability
\eq
P_k({\bf{b}},t\rightarrow \infty) = |a_k(t\rightarrow \infty)|^2.
\eqe

The coupled system of Eq. (\ref{mivan}) is solved numerically
by using a Runge-Kutta-Fehlberg method of fifth order with embedded
 automatic time step regulation. The conservation of the norm of the
 wave function is fulfilled better than 10$^{-8}$ during the collision.

      The eigenfunctions $\Phi_{j}$ in Eq. (\ref{TISE}) are obtained by
      diagonalizing the Hamiltonian
      in a basis of orthogonal
       symmetrized two-particle functions
       $f_{\mu}$ so that
       \eq
       \Phi_j({\bf{r}}_1,{\bf{r}}_2) = \sum\limits_{\mu}
       b_{\mu}^{[j]}f_{\mu}({\bf{r}}_1,{\bf{r}}_2)\,.
       \label{he:func}
       \eqe
In the applied independent particle model the many-particle wave functions $f_{\mu}$ are built up from single-particle orbitals. 
       For these single-particle wave functions,  an angular momentum 
 representation with spherical harmonics $Y_{l,m}$, hydrogen-like
       radial Slater functions and radial regular Coulomb wave packets is used.
       The Slater function can be written 
       \eq
       S_{n,l,m,{\kappa}}({\bf{r}}) = c(n,{\kappa})r^{n-1}e^{-{\kappa}r}Y_{l,m}
       ({\theta},{\varphi}),   \label{sl}
       \eqe
       where $c(n,\kappa)$ is the normalization constant.
        A regular Coulomb wave packet
 \eq
          C_{k,l,m,Z}({\bf{r}}) = q(k,\Delta k)Y_{l,m}({\theta},{\varphi})
              \int\limits_{E_k-\Delta E_k/2 }^{E_k+\Delta E_k/2}
              F_{k,l,Z}(r)\, dk           \label{pak}
              \eqe
              with normalization constant $q(k,\Delta k)$
 is constructed from the radial Coulomb function
              \begin{eqnarray}
                F_{k,l,Z}(r) = & \sqrt{{{2k}\over{\pi}}}
                 e^{{\pi\eta}\over{2}} {{(2\rho)^l}\over{(2l+1)!}} e^{-i\rho}
                  \mid \Gamma(l+1-i\eta)| \nonumber \times \\
                  &  _{1}\! F_{1}
                   (1+l+i\eta,2l+2,2i\rho),
                   \end{eqnarray}
                   where $ \eta = Z/k  $,  $ \rho = kr $.

The wave packets cover a small energy interval $\Delta E_{k}$
 and thereby forms a discrete representation of the continuum
 which can be incorporated into the finite basis set.
   The normalized Coulomb wave packets are calculated up to more than 
 300 a.u. radial distance to achieve a deviation of less than one percent
 from unity in their norm.

  In this approach, two different effective charges $Z$ for the target nucleus
 have been used to take into account the difference between the the singular and the double ionized stats of the He atoms. 
For single-ionized states, $Z=1$ and for the double-ionized case  $Z=2$  were used, respectively.  

The single- and double-continuum electron sates were calculated up to 6 a.u. energy equidistantly.

We  include  single-particle  wave  functions  with $ 0  \le l_1,l_2 \le   2$  angular momenta and couple them to 
$ 0 \le L \le 2  $ total angular  momentum  two-electron  states.  For  the $ L
= 0 $ configurations we use ss+pp+dd
angular correlated wave functions,
for $L = 1 $ we  use  sp+pd 
couplings  and  for $ L= 2$  the sd+pp+dd 
configurations, respectively.
For the ground state energy of He (1s1s)  the -2.901 a.u. was obtained which is reasonably
 accurate compared to the ``exact'' value of -2.903 a.u.

The diagonalization process gives 465 basis states which correspond to 1490 different collision channels, including 
different $m_l$ sub-states. The highest energy eigenvalue lies at 27.8 a.u. 
Numerous different basis sets were applied in order to test for convergence of the expansion (\ref{tim-w}).
 The results demonstrate that the channels with energies above 5 a.u  contribute very little to the
ionization probabilities.
The basis between the first ionization threshold (-2.0 a.u.) and the
lowest auto-ionizing  quasi-bound state ( ``2s2s"  E = -0.77 a.u. L=0), 
contained 22 discretized continuum states per total angular momenta, providing  the major contribution  for single-ionization.

The Feshbach projection \cite{imre} together with the complex scaling were adopted in order to separate excitation, 
double-  and single-ionization cross sections.
In the first step, a new ``reference'' Hilbert subspace was constructed and 
split into three different orthogonal subspaces characterized by the
properties of the two electrons: 1 - bound-bound, 2 - bound-ionized and 3 - ionized-ionized
electrons, respectively. In the second step, the 
``calculated Hilbert'' space was projected onto the reference space and this determined the excitation, single- and
double-ionization contributions.
To fix the effective charge of the Coulomb wave function used in the helium wave functions, 
the excitation and single-ionization cross sections were compared with the results obtained from
 the complex scaling \cite{imre4}.
Doubly-excited states embedded in the continuum e.g., ``2s2s''  (this labeling should not be taken literally because
of the strong electron-electron correlation \cite{ho}), can be  identified by the method of complex scaling and therefore the double-excitation
 and the single-ionization states can be separated. This new combination of the two methods is still not exactly rigorous but is much more feasible 
than the Feshbach method alone  and therefore reduces ambiguity.

To calculate the angular-differential ionization cross sections, the density operator was taken.   
The electron final-state density can be determined  from the time-depen-dent  wave function after the 
collision ($t \rightarrow \infty$) according to the expectation
value of the density operator $\hat{\rho} = \sum_{i=1,2} \delta
 ({\bf{r}}-{\bf{r}}_i) $
for a fixed impact parameter {\bf{b}}
\begin{equation}
\rho_{{\bf{b}}}({\bf{r}}) = \left\langle \Psi({\bf{r}}_1,{\bf{r}}_2
 ) \left | \sum_{i=1,2} \delta({\bf{r}}-{\bf{r}}_i)
 \right |  \Psi({\bf{r}}_1,{\bf{r}}_2
  )\right \rangle. 
  \end{equation}
In order to extract the angular distribution of the
ionized electrons two additional operations are needed: 
\begin{enumerate}

\item The wave function $\Psi$ is projected onto the single-ioni-zation
continuum  $|\Psi_{ion}  \rangle  = (1- \hat{P}_b - \hat{P}_{di}) \> \> |
 \Psi \rangle $ where  $\hat{P}_b $ is the projector
onto the bound state subspace (including all excited states) and  $\hat{P}_{di} $ is the projector
onto double-ionized states.
\item The radial and the azimuthal coordinates have to be integrated over
to get the polar angle distribution of the ionized electrons:
\eq
P_{{\bf{b}}}(\theta)= \frac{1}{2\pi}\int\limits_{0}^{2\pi} \int\limits_{0}^{\infty}
    \langle\Psi_{ion} | \sum_{i=1,2} \delta({\bf{r}}-
{\bf{r}}_i) |  \Psi_{ion}\rangle r^2drd\varphi
    =\nonumber  
    \eqe
    \eq
  \frac{1}{\pi}\int\limits_{0}^{2\pi}\int\limits_{0}^{\infty}
    \int\limits_{{\bf{r_1}}}\left|
         \Psi_{ion}    \right|^2
       d^3r_1r^2drd\varphi. 
\eqe
\end{enumerate}
The angular differential cross section is obtained by integrating
$P_{\bf{b}}(\theta)$ over the impact parameter. 
This method has  already been used in a previous publication \cite{barna0}  and it gives a
satisfactory agreement with other theoretical results. 

Contrary to other perturbative or classical approaches, TDCC methods  has a finite number of discretized final-states.
This is why it is not possible  to calculate energy-differential cross sections in a rigorous way -- only the distribution of the channel cross sections divided 
by an effective energy can be obtained This effective energy can be defined in different ways, it can be the energy 
of the corresponding channel; the difference of the energies of the 
neighboring two channels or the widths of the Coulomb wave packets (\ref{pak}). 
Here we consider the last one, and we use the form for calculating the approximated energy differential cross sections according to the following equation:
\eq
 \frac{d \sigma}{d E} \approx  \frac{\sigma_k}{\Delta E_k}.
\eqe

\section{Results and discussions}
\label{Results}
Table 1 shows our recent total single-ionization cross sections with some of the latest theoretical data.
Figs. 1a, 2a, 3a  display the DDCS for single ionization of helium at 10, 50 and 100 keV antiproton impact energies 
within the framework of the TDCC method.
According to the figures, generally, we can conclude the followings:
In each cases we found a minimum at zero angle indicating the existence of the Coulomb hole (``anti-cusp'').  
In general we can say, that the anti-cusps are always properly appear with antiproton projectile impact. As the projectile energy enhances the centers of the anti-cusps enhance too. At 10 keV the DDCS are generally smaller than for 50 or 100 keV. Despite the numerous theoretical studies investigations into the anti-cusp \cite{burg,26,28} over the last decades, there is still no experimental observation of the anti-cusp
in anti-proton impacts. Slow electrons are possibly ejected into the backward direction and the fast 
electrons are ejected along the broad ridge identified  
as the binary encounter (BE) ridge.  
\begin{table}
\begin{center}
\begin{tabular}{|r|r|r|r|}
\hline \hline\, Energy [keV ]& $\sigma^+$ (our)   &  $\sigma^+$ \cite{foster}   & $\sigma^+$   \cite{keim2} \\
\hline
  10 & 0.45  &  0.43  &  0.48 \\
\hline
 50 & 0.58  &    0.62  & 0.63  \\
\hline
 100 & 0.61  &    0.60 &  0.68\\
\hline
\hline
\end{tabular}
   
\caption{Single-ionization cross sections ($\sigma^+$)for the investigated energies. \label{tab:Ens} The second column presents our results, the third one 
is from \cite{foster} and the  fourth one is from \cite{keim2}. 
All numerical values should be multiplied by $10^{-16} $cm$^2$. } 
\end{center}
\end{table}
This BE ridge may be the easiest feature to understand as fast electrons are ejected 
through a series of hard collisions between the projectile and the target known as
the Fermi-shuttle mechanism \cite{30}.  
For all three energies (10, 50 and 100 keV), the ridge follows closely this trend.

The distribution of slow electrons in backward directions is a clear indication of
final state interactions between the ejected electron and the antiproton. As the
force between the electron and the anti-proton is repulsive, slow electrons $(v_e < v_p,$ where 
$v_e$ and $v_p$ are the electron and projectile speeds, respectively) lag behind the antiproton
and will be scattered to the backward direction. For the total cross section the contribution of slow electrons 
is dominated therefore the angular distribution in the DDCS will be suppressed in the forward direction, 
as clearly shown in Fig. 1a, 2a and 3a. This is in sharp contrast to proton impact, where single differential cross sections will generally peak at small angles and decrease toward large angles. The remarkable contrast to proton impact is the anti-cusp (void) region for anti-proton impact. This can also be attributed to the repulsive final state interactions.

To have a more transparent overview of the trends of our calculations we show additional figures, namely cuts at constant energies and cuts at constant scattering angles for all the three energies.

Figure 1b shows the energy differential cross sections at given angles. The cuts are parallel to the 
x (or energy) axis at 0, 60 and 120 degrees which are at the cusp, 
at the ridge and above the ridge.  
The presence of the anti-cusp is evident. At low electron energies, (around the ionization threshold) 
the cross sections are 10000 times lower for 
20 degree emission angle than for 60 degree emission angle.  
Note that a factor of 100 remains even at 70 eV  electron energies.  
The ratio of the cross sections between 60 and 120 degrees are much smaller.

Fig. 1c shows the angular differential cross sections at 20, 40 and 60 eV energies. 
We note that  at zero scatterting angle the electrons with 20 eV kinetic energy have a 10000 times 
smaller cross section than the electrons with energy of 40 eV. 

Fig. 2a shows the DDCS for 50 keV antiproton impact energy. 
The general features remain the same---however, all differential cross sections are larger in accordance with the total cross sections. 
The position of the anti-cusp lies at 30 eV.
On the contrary the maximal widths of the corresponding anti-cusp angle is reduced from 20 degrees to 5. Fig. 2b shows the cross sections at different emission angles. Note, that the cross sections lie in almost the same magnitude in all directions. 
This is drastically different to the 10 keV energy case. Fig. 2c shows the angular distributions
for 20 and 40 eV energy electrons. Here again at zero scatterting angle the electrons with 20 eV kinetic energy have a 10000 times smaller cross sections than the electrons with energy of 40 eV

Figure 3 shows the ionisation cross sections for 100 keV antiproton with the same general features. The center of the 
anti-cusp is at 50 eV with the energy widths of $\pm 30 $ eVs.  
It is clear to see in Fig. 3b that at zero emission angle and at backscattering all the three cross sections lie in almost the same magnitude. 
Fig. 3c shows a significant difference to Fig. 2c and Fig. 1c in that the cross section curves does not cross---a clear fingerprint that the 
whole distribution has become much more flat. 
\begin{figure}
\resizebox{0.7 \textwidth}{!}{
\includegraphics{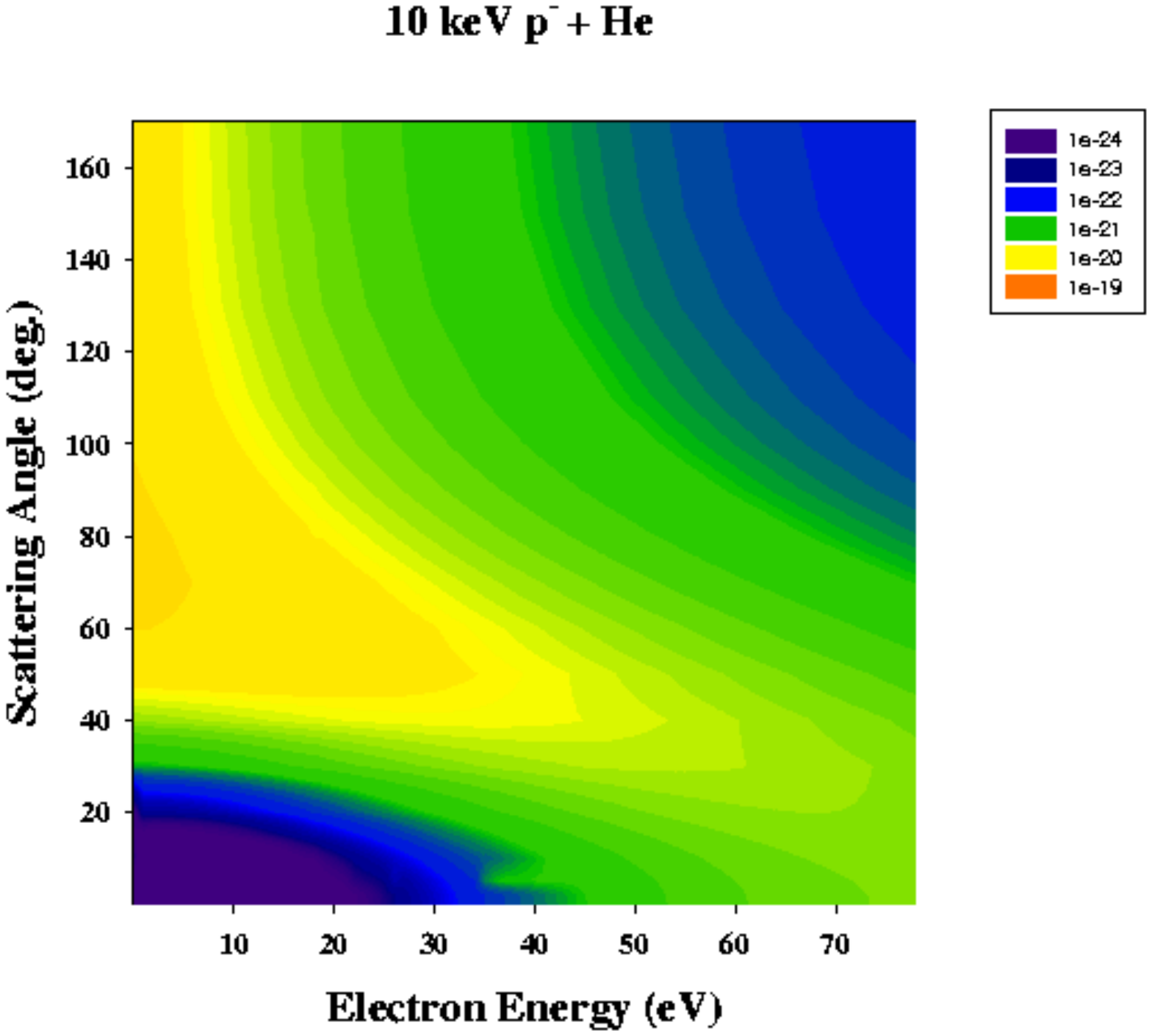}
} 
\label{fig:1}       
\resizebox{0.95\textwidth}{!}{
  \includegraphics{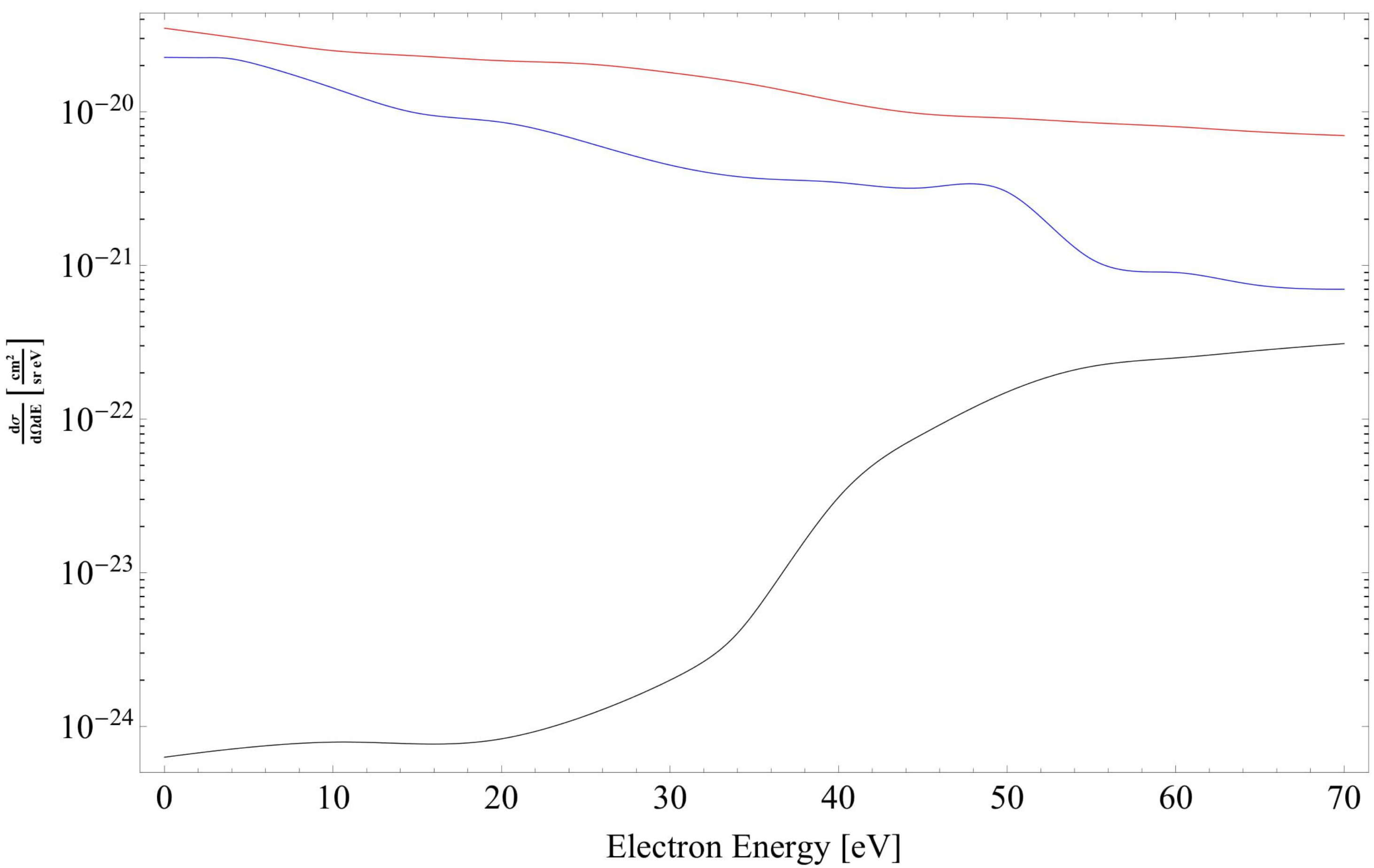}
 \includegraphics{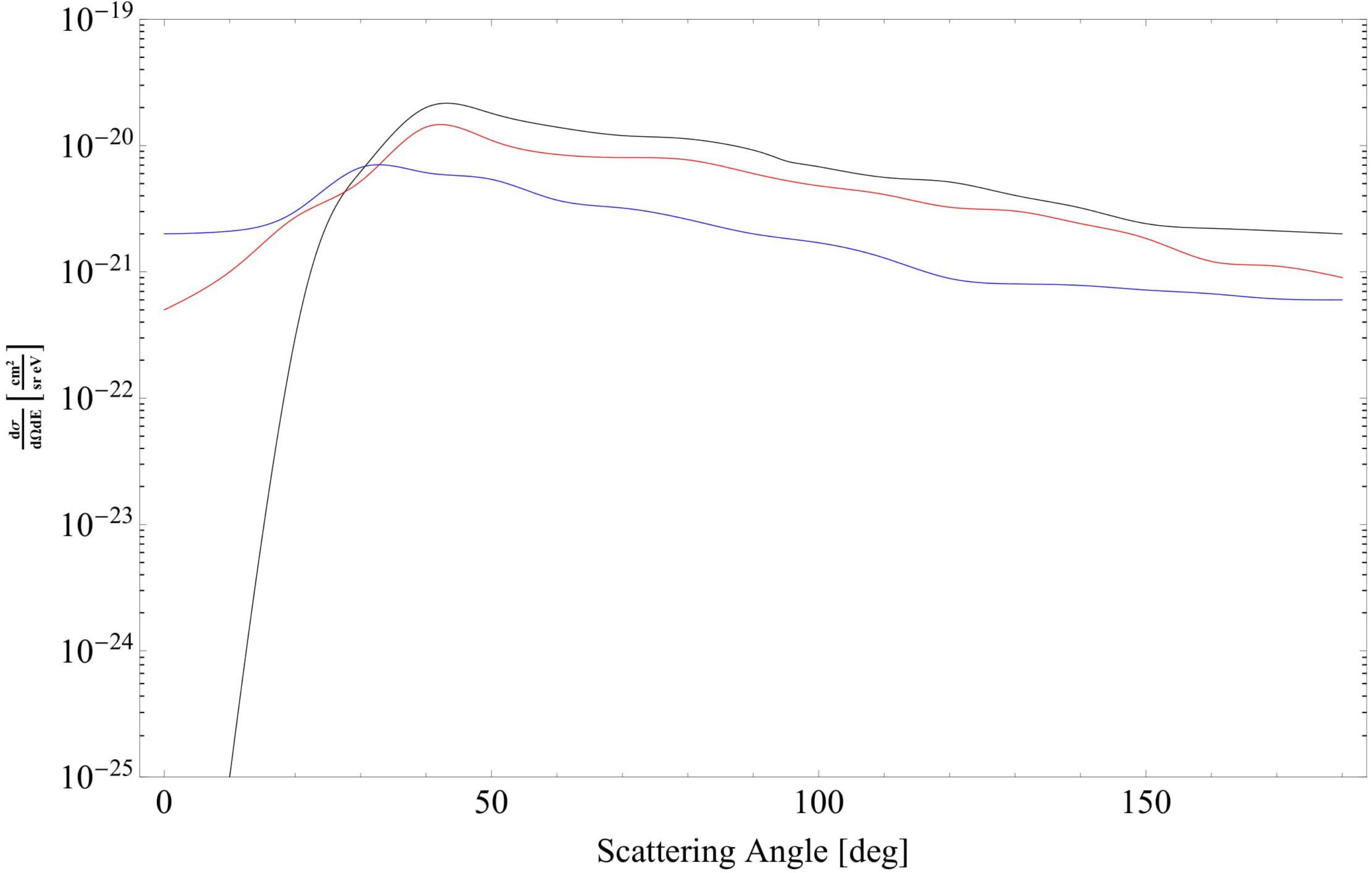}
}
\caption{ a) Doubly differential electron emission cross sections emitted from He and 10 keV antiproton impact.  \\
b) Three cuts of the DDCS at different angles. The black line is for 0, the red line is for 60 and the blue one is for 120 degrees. \\
c) Three cuts of the DDCS at different energies. The black line is for 20, the red line is for 40 and the blue one is for 60 eV.}
\label{fig:1c}       
\end{figure}
\begin{figure}
\resizebox{0.7\textwidth}{!}{%
  \includegraphics{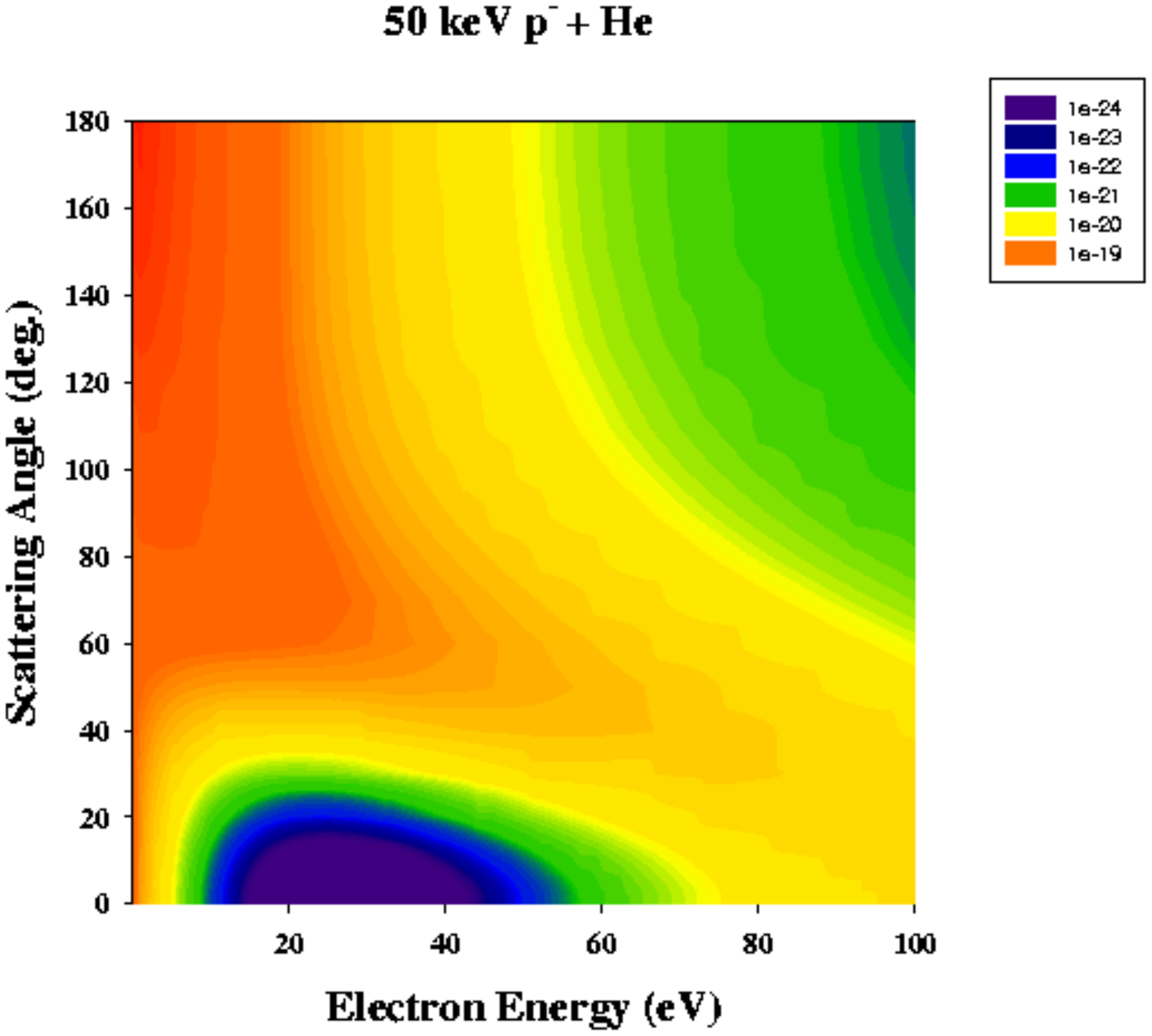}
}
\resizebox{0.95\textwidth}{!}{%
 \includegraphics{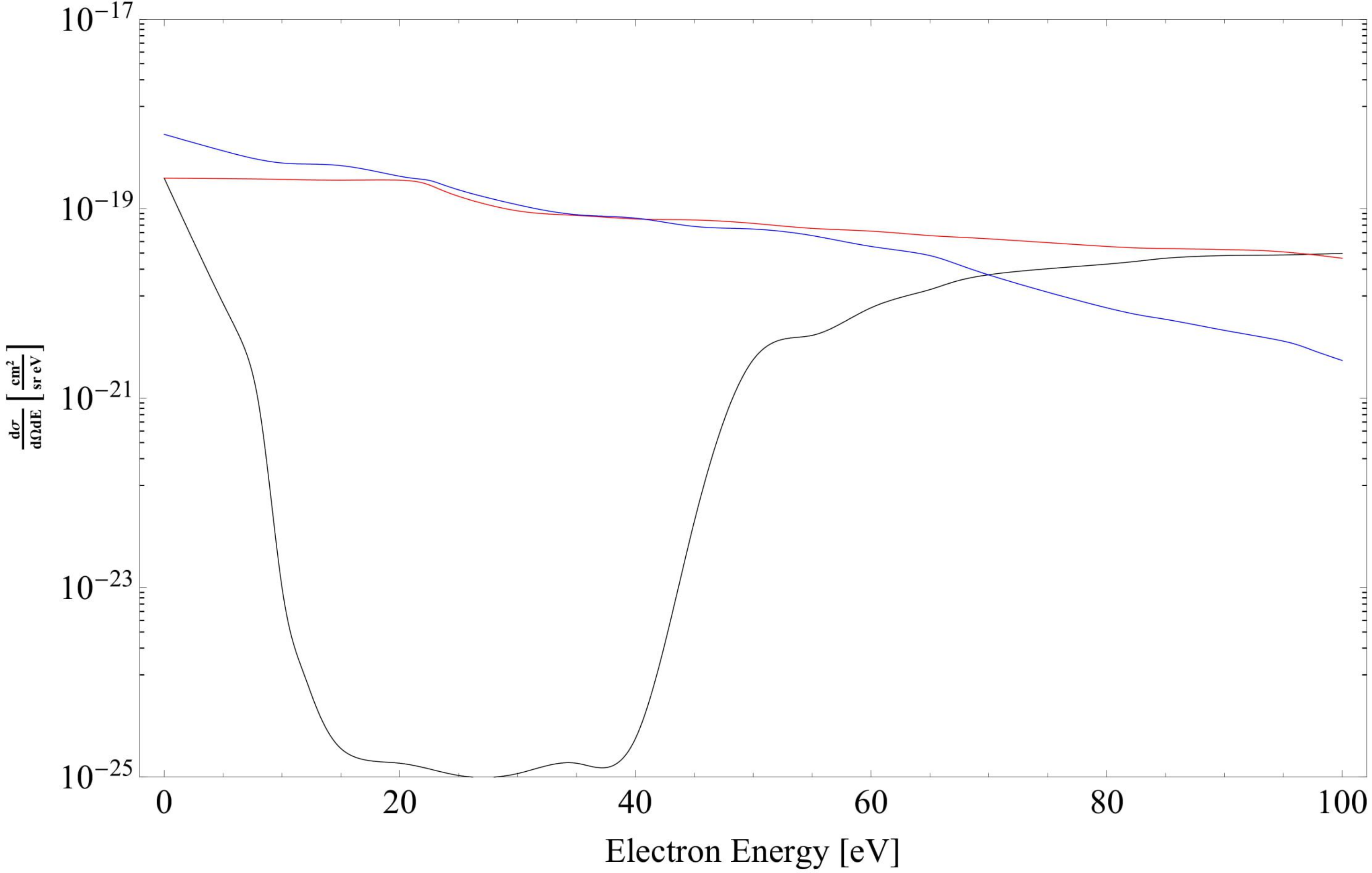}
  \includegraphics{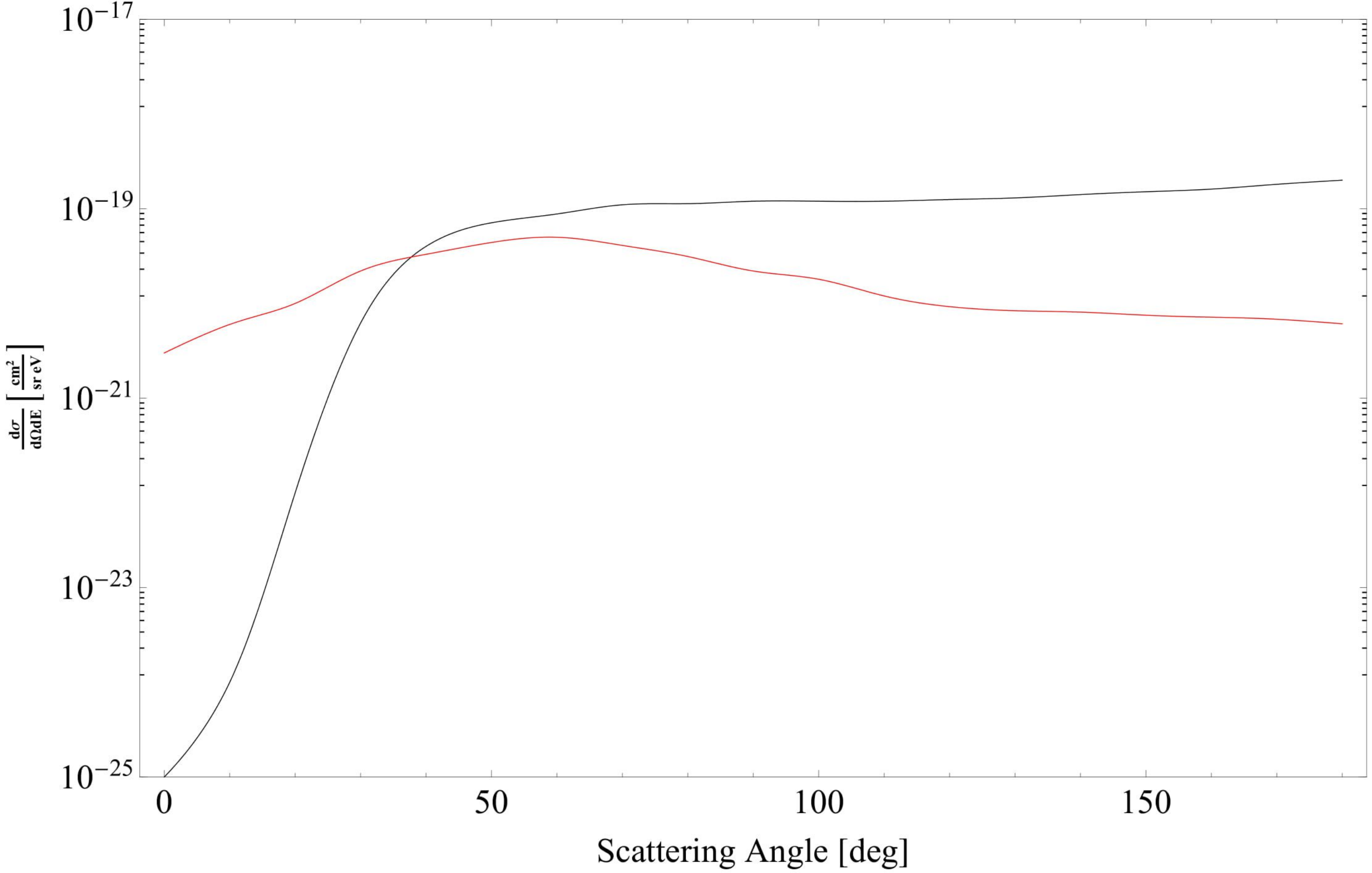}
}
\caption{ a) The same as Fig. 1 but for 50 keV impact energy. \\
b)Three cuts of the DDCS at different angles. The black line is for 0, the red line is for 60 and the blue one is for 120 degrees.  \\
c) Two cuts of the DDCS at different energies. The black line is for 20, the red line is for 40 eV.}
\label{fig:2c}       
\end{figure}
\begin{figure}
\resizebox{0.7\textwidth}{!}{%
  \includegraphics{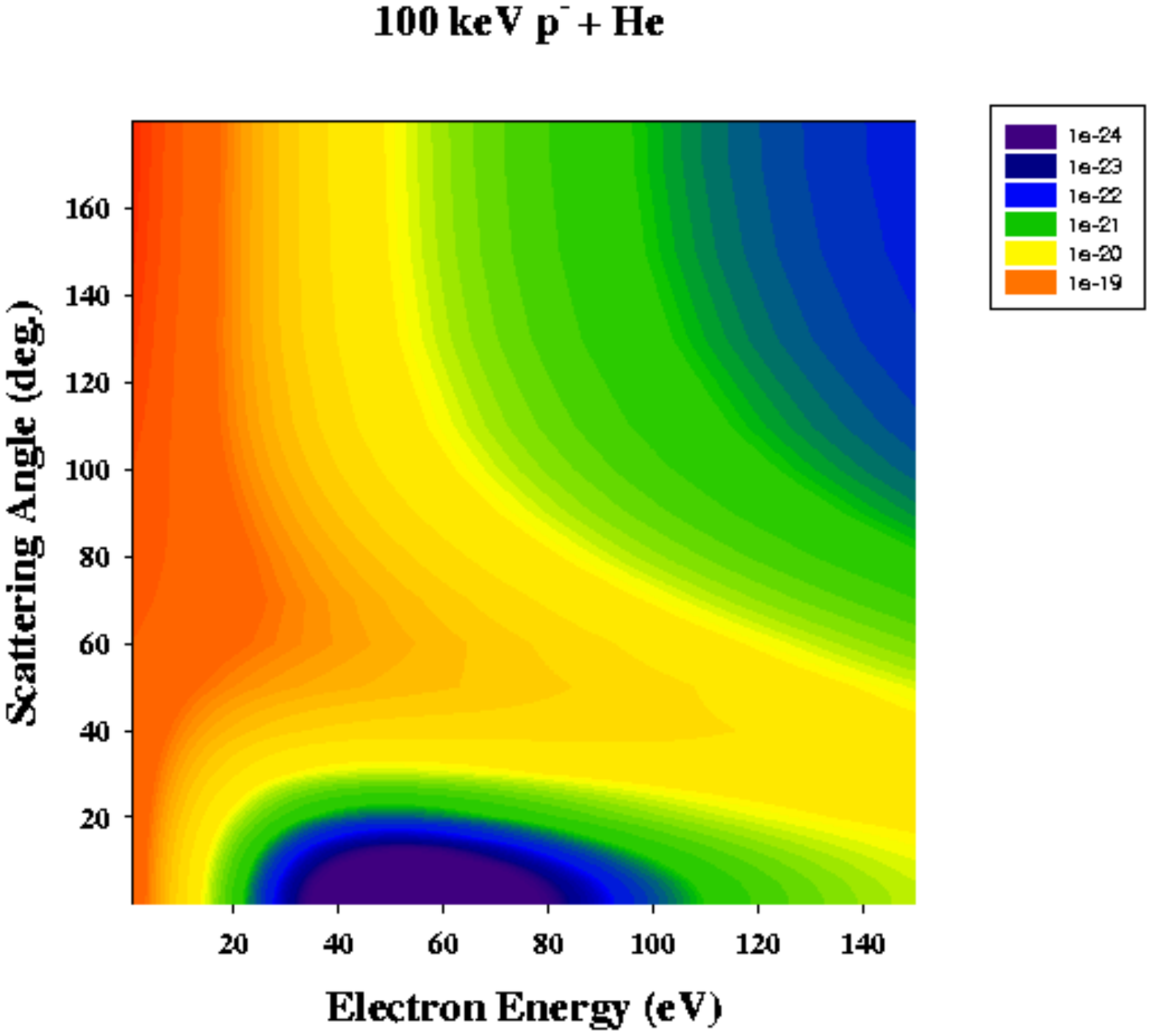}
}
\label{fig:3}       
\resizebox{0.99\textwidth}{!}{%
  \includegraphics{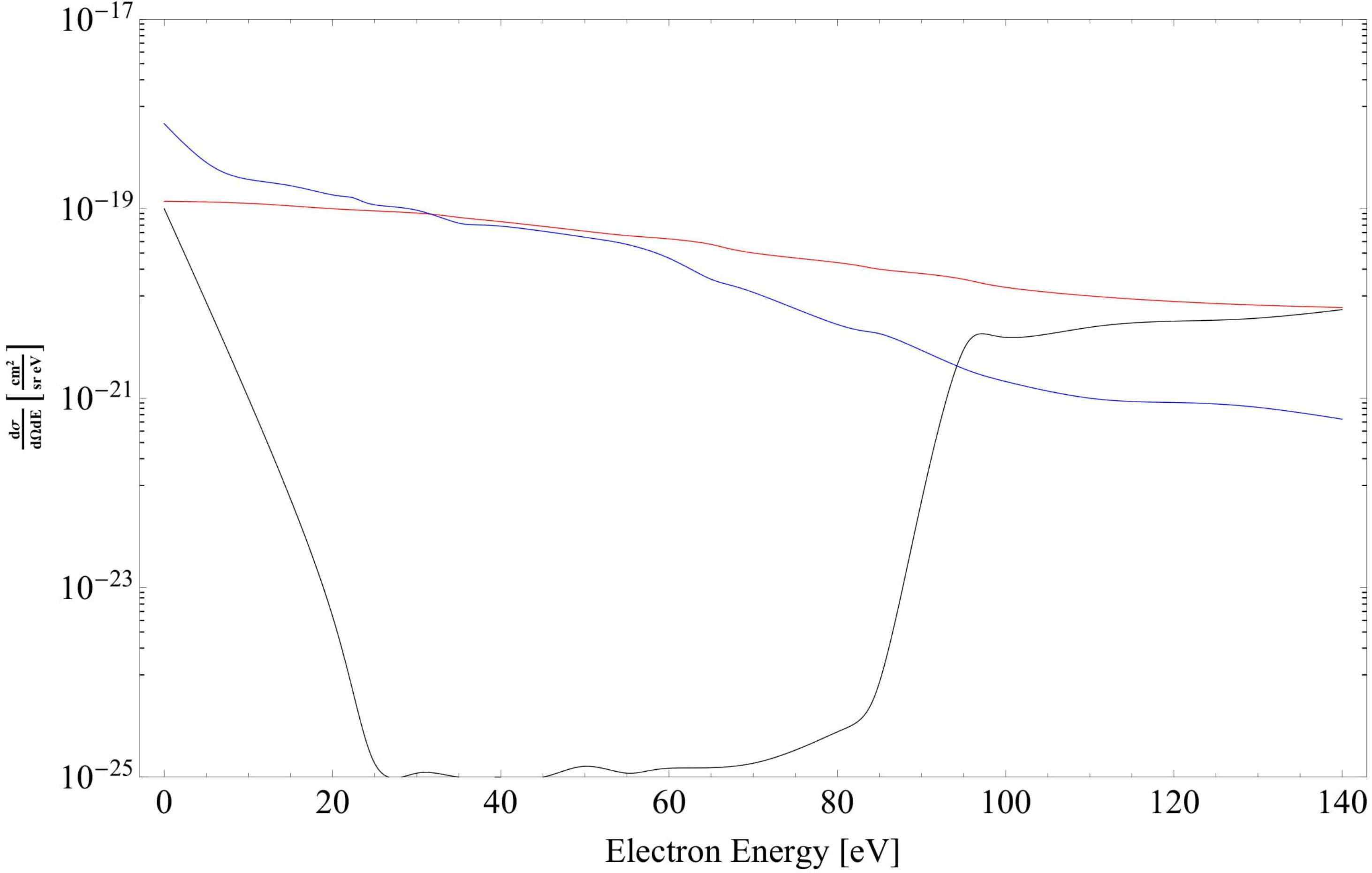}
 \includegraphics{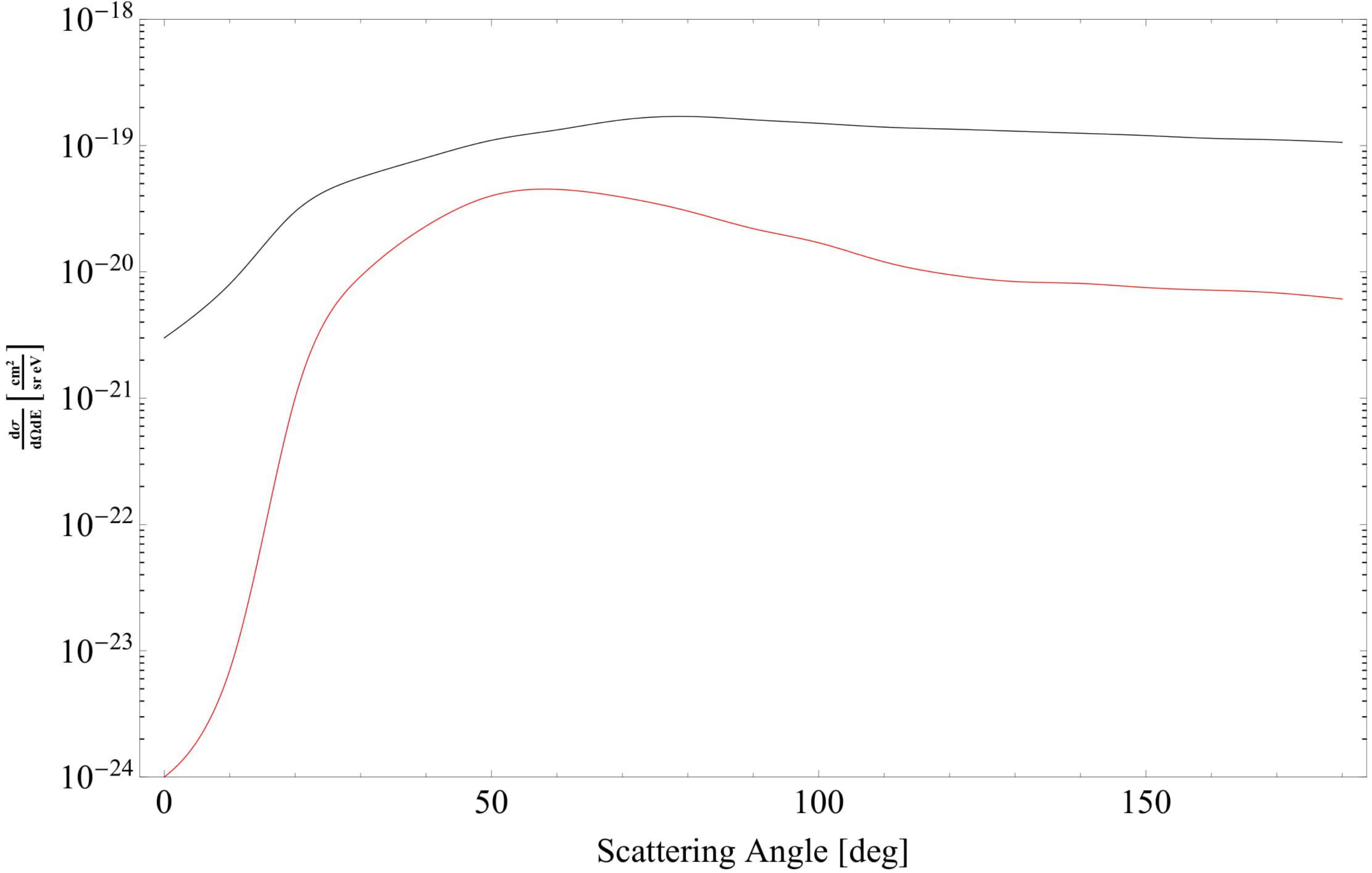}
}
\caption{a)The same as Fig. 1 but for 100 keV impact energy. \\  
b) Three cuts of the DDCS at different energies. The black line is for 30, the blue is for 60 and the red one is for 120 degrees.  \\
c)Two cuts of the DDCS at different energies. The black line is for 20, the red line is for 40 eV.}
\label{fig:3c}       
\end{figure}

\section{Summary and conclusions}
The singly and doubly differential ionization cross sections for antiproton and helium atom collisions at 10, 50 and 100 keV impact energies were presented. The calculations were based on {\it{ab-initio}} time-dependent coupled channel method using a finite basis set of regular helium Coulomb wave packets and Slater function. A semiclassical approximation was applied with the time-dependent Coulomb potential to describe the antiproton electron interaction. 
We found a strong final state interactions between the antiprotons and the ejected electrons in the forward scattering angles. We clearly identified the existence of the formation of anti-cusp for each antiproton impacts. We hope that our recent calculations will further encourage the experimentalist to carry out differential cross section measurements in the in the near future.
 
\section{Acknowledgments}
We would like to thank Dr. Andrew Cheesman for the critical reading of the manuscript.
The ELI-ALPS project (GINOP-2.3.6-15-2015-00001) is supported by the European Union and co-financed by the
European Regional Development Fund. 
One of us (KT) was also supported by the National Research, Development and Innovation Office (NKFIH) Grant KH126886, 
and by the European Cost Actions CA15107 (MultiComp). 

\end{document}